\documentstyle[prl,twocolumn,aps,epsfig]{revtex}

\newcommand{\be}{\begin{eqnarray}}
\newcommand{\ee}{\end{eqnarray}}

\newcommand{\ket}[1]{\left| #1 \right\rangle}

\begin{document}

\title{
Observation of coherent transients in ultrashort chirped
excitation of an undamped two-level system. }
\author{S\'ebastien Zamith, J\'er\^ome Degert, Sabine Stock, B\'eatrice de Beauvoir,
Val\'erie Blanchet, M. Aziz Bouchene and Bertrand Girard}
\address{
Laboratoire de Collisions, Agr\'egats et Reactivit\'e (CNRS UMR 5589), IRSAMC\\
Universit\'e Paul Sabatier,
31062 Toulouse, France}
\date{\today}
\maketitle

\begin{abstract}

 The effects of Coherent excitation of a two level system with a linearly
 chirped pulse are studied theoretically and experimentally (in Rb (5s - 5p))
 in the low field regime.
The Coherent Transients are measured directly on the excited
state population on an ultrashort time scale. A sharp step
corresponds to the passage through resonance. It is followed by
oscillations resulting from interferences between off-resonant and
resonant contributions. We finally show the equivalence between
this experiment and Fresnel diffraction by a sharp edge.

\end{abstract}


Coherent excitation of atomic transitions is a general phenomenon
occurring whenever electromagnetic radiation interacts with atoms.
This interaction may result in various kinds of processes, for
example Rabi oscillations, free induction decay, adiabatic
population transfer and Coherent Transients (CT)  \cite{AllenEberly74}. By simply varying the temporal shape of the pulses, a
great variety of systems can be manipulated such as spins in
Nuclear Magnetic Resonance  \cite{Abragam61} or atomic and
molecular systems in coherent control schemes  \cite{Judson92}. We consider here the simplest case of linearly chirped
pulses where the ''instantaneous frequency'' drifts in time.

Depending on the frequency sweep and on the intensity, adiabatic
following can be observed with a significant population transfer.
For example total population transfer has been achieved via
multilevel ladder climbing  \cite{Eberly85,CorkumChirp90,WarrenI291,Noordamladder92,Band94b,NoordamNO98} and Stimulated Raman Adiabatic Passage
 \cite{Bergmann90}.

On the contrary if the adiabaticity criterion is not fulfilled,
the final population depends crucially on the pulse integral and
CT dominate the interaction. This letter presents theoretical and
experimental studies of these Coherent Transients. Although many
works have dealt with the transfer efficiency into the final
state, we report here the first direct observation of these
transients on a subpicosecond time scale. The CT result in
oscillating population. This is a signature of interferences
between resonant and non-resonant excitation. A detailed
understanding of this behaviour is crucial to analyze pump-probe
experiments in which dynamics takes place on the same time scale
as laser interaction. Due to the difficulty to control properly
the laser pulses, the effects of CT are most often ignored
 \cite{Trushin00}.

Similarities can be found with optical CT observed in the
experiments of free induction decay or photon echoes which provide
relaxation rates  \cite{Brewer71,Glorieux72,Grischkowsky76}. In these experiments, they observed
on the transmitted optical signal, a beat between initially
induced polarisation of the medium at resonance and a
near-resonant laser. In the particular case of a single chirped
optical pulse, the beat frequency increases linearly after
passage through resonance, following the increase of resonance
mismatch  \cite{Grischkowsky86}. On the contrary to these
previous studies, the present report involves CT directly
observed on the population before relaxation becomes significant.
The beats result here from interferences between the resonant and
non-resonance excitation paths. The passage through resonance
does not lead to total inversion  \cite{Vitanov99}. Thus
non-resonant excitation can lead to population transfer after
resonance. However, these extra contributions have a phase which
varies rapidly with respect to the phase of the dominant
contribution, leading to population oscillations.

In this letter, we describe theoretically the temporal evolution
of the excited state population induced by chirped pulse
excitation in the low field regime. The CT oscillating pattern is
experimentally studied on the Rb (5s $\rightarrow$ 5p)
transition. The 5p population is probed "in real time" by a
second excitation induced by an ultrashort pulse overlapping
partially with the excitation pulse. These results illustrate the
relative importance of the various stages of the interaction.
Most of the population transfer occurs at resonance. The small
fraction of excited state amplitude transferred after resonance
leads to strong interferences, whereas interaction before
resonance results in negligible effects. In order to point out
the different stages of the interaction, two cases are
investigated, the first being when resonance is reached at the
middle of the interaction (on-resonance case) and secondly when
the resonance is reached earlier (off-resonance case).

\begin{figure}[!ht]
\begin{center}
\epsfxsize 5cm \epsfbox{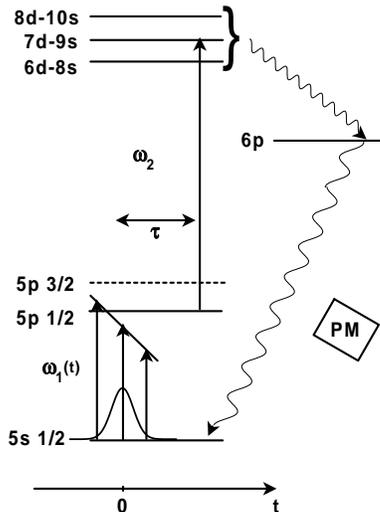} \caption[principe]
{Simplified energy-level diagram of Rubidium. The pump pulse is
chirped (and drawn here off-resonance).} \label{principe}
\end{center}
\end{figure}

The excitation scheme is depicted on Fig. \ref{principe}. A
quantum system with three states $\ket{g}$, $\ket{e}$ and
$\ket{f}$, of increasing energies $\hbar \omega_g$, $\hbar \omega_e$ and
$\hbar \omega_f$ respectively, interacts with two ultrashort laser
pulses $E_1 (t)$ and $E_2(t)$. These laser pulses have carrier
frequencies $\omega_1$ and $\omega_2$, each one close to resonance
with an atomic transition : $\omega_{eg} =\omega_e - \omega_g
\simeq \omega_1 $ and $\omega_{fe} =\omega_f - \omega_e \simeq
\omega_2$. Moreover, these transitions
 are sufficiently separated compared to the laser
linewidths ($\left| \omega_{fe} - \omega_{eg} \right| \gg \Delta
\omega_1, \Delta \omega_2$) so that each laser can be assumed to
interact with one atomic transition only.

The first laser pulse is obtained by frequency chirping a Fourier
transform limited laser pulse of duration $\tau_1 = \frac{4 \ln 2}{\Delta \omega_1}$.
 The second one is kept as short as possible in
order to probe the transient population. Keeping only the resonant
component (within the rotating wave approximation), the electric
fields of the pulses are given by (for $n=1,2$) $ E_n \left( t
\right) = E_n (0)\exp \left( { - \Gamma _n t^2 } \right)\exp
\left( {-i\omega _n t} \right) $
 where $ \Gamma _n = \left[
{\frac{{8\,\,\ln 2}}{{\Delta \omega _n ^2 }} - 2i\phi ''_n}
\right]^{ - 1}$ and the new pulse duration is $ \tau _{nc}  = \tau
_n \sqrt {1 + \frac{{\left( {4 \ln 2 \,\phi ''_n} \right)^2
}}{{\tau _n^4 }}} $. Here $\phi ''_n= \frac{{d^2 \phi _n
}}{{d\omega ^2 }} $ is the quadratic dispersion responsible for
the Group Velocity Dispersion induced in the pulse by the
chirping process. In this letter, we concentrate on the case
where the pump pulse is strongly chirped, so that $ \tau _{1c}
\simeq {{2 \phi''_1} \mathord{\left/
 {\vphantom {{2 \phi''_1} {\tau _1}}} \right.
 \kern-\nulldelimiterspace} {\tau _1}} $.
Moreover, to measure the "instantaneous" excited state population (in
state $\ket{e}$), the probe pulse is assumed to be nearly Fourier
transform limited ($\phi'' _2 \simeq 0$), and of short duration
compared to the chirped pump pulse duration ($\tau _2 \ll \tau_{1c}$).
This will ensure that the final state population (in state $\ket{f}$)
reproduces the $\ket{e}$ state population.

In the low field regime, analytic solutions of the Schrödinger
equation
are easily found within first order perturbation theory, with a
probability amplitude of state $\ket{e}$ proportional to:

\be \label{amplitude e}
 a_e (t) \propto \int_{
- \infty }^t {\exp \left( { - \frac{{2 \ln 2 t'^2 }}{{\tau _{1c}^2 }}}
\right)} \,\,\exp \left( { - i\frac{{\left( {t' - t_0 }
\right)^2  + t_0^2 }}{{2\phi_1 ''}}} \right)dt' \ee where $ t_0 =
\left( {\omega _{eg}  - \omega _1 } \right)\phi_1 '' $ is the time
when the laser sweeping frequency goes through the atomic
resonance.

In the stationary phase approximation, the main
contribution to $ a_e \left( t \right)$ arises from delays within
the interval $ \left[ {t_0  - \sqrt {\phi_1 ''} ;\,t_0  + \sqrt
{\phi_1 ''} } \right] $. This corresponds to the interval during
which the laser frequency is resonant with the atomic transition.
Non-resonant transitions can also contribute to the excited state
amplitude, but with a smaller efficiency. As time increases from
$t_0$, new contributions are added to the amplitude $ a_e \left(
t \right)$. These contributions have a phase which increases more
and more rapidly, leading alternatively to an increase and a
decrease of the excited state amplitude.



In order to illustrate the coherent transients, an experiment has
been performed in Rubidium atoms. The (5s - 5p ($^2 {\rm
P}_{1/2}$)) transition (at $795 \, {\rm nm}$) is excited either
with a Fourier transform limited pulse or with a chirped pulse.
The laser bandwidth of $\sim 10 \, {\rm nm}$ limits the excitation
of the other fine structure component 5p ($^2 {\rm P}_{3/2}$))
transition (at $780 \, {\rm nm}$). The transient excited state
population is probed on the (5p - ns,n'd) transitions with an
ultrashort pulse (at $607 \, {\rm nm}$).


The laser system is based on a conventional Ti: Sapphire laser
with chirped pulse amplification (Spitfire Spectra Physics)
producing $130 \, {\rm fs}$ pulses around $795 \, {\rm nm}$ at a
$1 \, {\rm kHz}$ repetition rate with energy of as much as $ 800
\, {\rm \mu J}$. Half of the light is used as the pump pulse. The
other half feeds a home made Non-collinear Optical Parametric
Amplifier (NOPA)  \cite{DeSilvestri97} compressed with Brewster-cut prisms,
which delivers pulses of a few microJoules, $30  \, {\rm fs}$
(after compression), centered around $607  \, {\rm nm}$. The pump
pulse is negatively chirped with a pair of gratings ($ \left|
\phi_1 '' \right | \gtrsim 2.\, 10^5 \, {\rm fs}^2$)), recombined
with the probe pulse and sent into a sealed rubidium cell with
fused silica Brewster-window ends ($3 \, {\rm mm}$ length). The
cold finger of the cell is maintained at about $ 80 \,^\circ {\rm
C}$, corresponding to a pressure of $4. \, 10^{-5} \, {\rm
mbar}$. The pump-probe signal is detected by monitoring the
fluorescence at $420 \, {\rm nm}$ due to the radiative cascade
(ns,n'd) $\rightarrow$ 6p $\rightarrow$ 5s collected by a
photomultiplier.

\begin{figure}[!ht]
\begin{center}
\epsfxsize 8cm \epsfbox{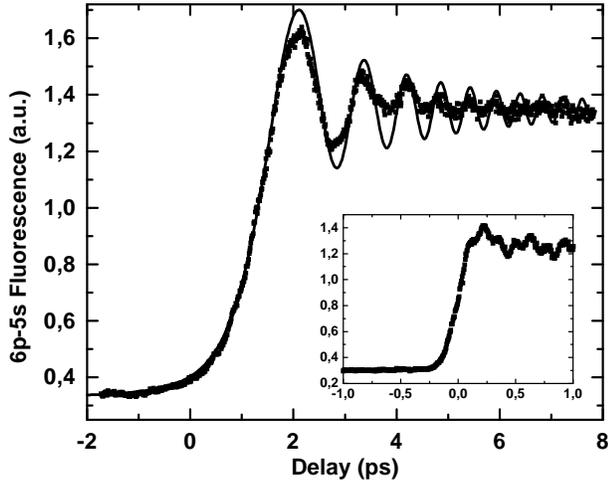} \caption[resonance]
{Experimental results and numerical solution obtained on resonance ($\lambda_1 = 795 \,
{\rm nm}$) for a chirp of $ \phi_1 ''
 = -4.\, 10^5 \, {\rm fs}^2$ and for a chirp-free pulse in the inset.} \label{resonance}
\end{center}
\end{figure}

\begin{figure}[!ht]
\begin{center}
\epsfxsize 8cm \epsfbox{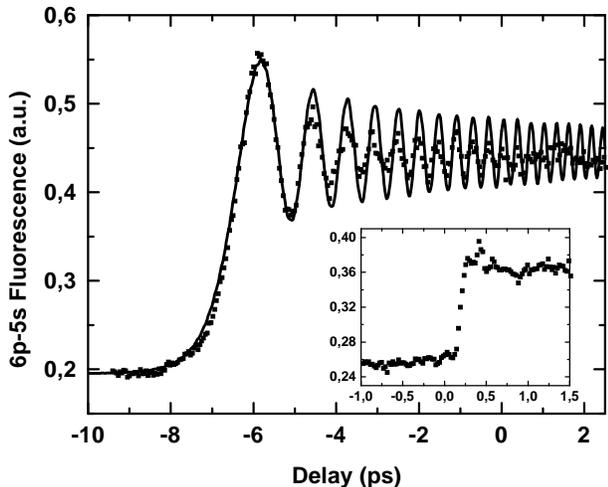} \caption[belowresonance]
{Experimental results and numerical solution obtained below resonance ($\lambda_1 = 801 \,
{\rm nm}$) for a chirp of $ \phi_1 ''
 = -4.\, 10^5 \, {\rm fs}^2$ and for a chirp-free pulse in the inset} \label{belowresonance}
\end{center}
\end{figure}

Figs. \ref{resonance} and \ref{belowresonance} present the
experimental pump-probe signals. Measurements have been performed
for resonant excitation ($\lambda_1 = 795 \, {\rm nm}$ on Fig.
\ref{resonance}) and for a central wavelength detuned below
resonance by roughly half the laser bandwidth ($\lambda_1 \simeq
801 \, {\rm nm}$ on Fig. \ref{belowresonance}). Each figure
presents a fluorescence signal obtained with a chirp of $ \phi_1
''
 = -4.\, 10^5 \, {\rm fs}^2$ corresponding to a pulse duration of $11 \, {\rm ps}$.
As a comparison, chirp-free data are displayed in the insets.
In this latter case, the pump-probe signals show a sharp step,
with a short rise time corresponding to the laser
cross-correlation duration. On-resonance, the plateau is slightly modulated
by spin wavepacket dynamics with a period of $140 \, {\rm fs}$.
Indeed, off-resonance population of the 5p $^2 {\rm P}_{3/2}$
state is induced by the spectral tails of the pulse. The coherent
superposition of the two excited states corresponds to a fine
structure wave packet which oscillates  \cite{Jones95MultiEnhance,Zamith2000a}. These oscillations are well-known and
not related to the present study.

With a significant chirp, we observe large amplitude oscillations which
are the signature of the Coherent Transients. The resonance experiment
(Fig. \ref{resonance}) was performed in the perturbative regime
(fluence of $ 2 \, \mu {\rm J/cm}^2$). Due to the smaller efficiency of
the off-resonance excitation ($\lambda = 801 \, {\rm nm}$, Fig.
\ref{belowresonance}), the fluence was increased to $110 \, \mu {\rm
J/cm}^2$, inducing some power effects. However, experiments performed
for different fluences show qualitatively the same behaviour.

\vspace{-1cm}

In each case, the experimental curve is compared with the
numerical solution of the Schr\"{o}dinger equation projected on
the 5s, 5p ($^2 {\rm P}_{1/2}$, $^2 {\rm P}_{3/2}$), 6d up to 10s
states. The comparison with the numerical resolution (where this
time origin is well defined) has been used to adjust the relative
time origins of the various experimental data. Indeed, in the
case of long chirped pulse, the experimental determination of the
time delay origin is a real challenge via standard setup based on
frequency mixing. The CT scheme could provide an alternate
experimental determination of both time origin and long linear
chirp measurements. The only fitted parameter is the value of the
chirp which differs by less than $2 \%$ from the estimates
deduced from the geometry of the stretcher. An excellent
agreement on the oscillation pattern is obtained. The contrast of
the first periods is well reproduced. It drops more rapidly in
the experiment. Several causes can be invoked : residual spatial
chirps or small incoherent components of the pump pulse. After
the sharp step, the oscillation frequency of the interferences
increases linearly with time : $ \omega_{eg} - \omega_1\left( t
\right) = {t \mathord{\left/ {\vphantom {t {\phi ''_1 }}} \right.
\kern-\nulldelimiterspace} {\phi ''_1 }} $. For a pulse frequency
below resonance and a negative chirp, the passage through
resonance occurs at the beginning of the pulse. Thus the major
part of the pump spectral components contributes after the
resonance so that many oscillations result as can be seen on Fig.
\ref{belowresonance}. On the contrary, in the resonant case, only
the second half of the pulse contributes to the interferences and
fewer oscillations are observed.

\begin{figure}[!ht]
\begin{center}
\epsfxsize 8cm \epsfbox{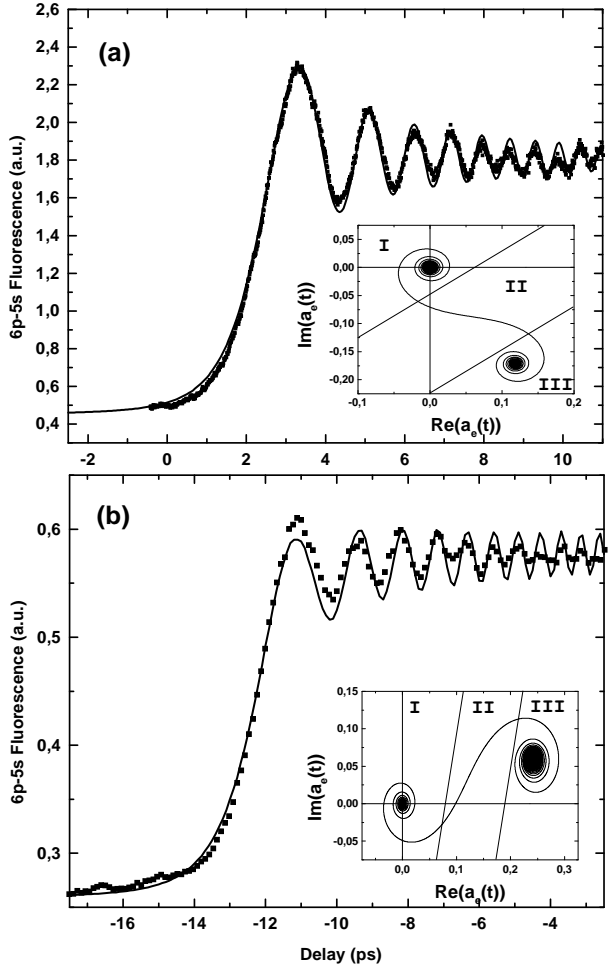} \caption[Cornu]
{Experimental results obtained a) on resonance ($\lambda_1 = 795 \,
{\rm nm}$), b) below resonance ($\lambda_1 = 801 \, {\rm nm}$) for
a chirp of $-8. \, 10^5 \, {\rm fs}^2$ and the corresponding
excited state amplitude (numerical resolution of the Schrödinger
equation) drawn in the complex plane (insets).} \label{Cornu}
\end{center}
\end{figure}

An other way to explain the CT phenomenon is to examine the
behaviour of $a_e \left( t \right)$ in the complex plane as
displayed in the insets of Fig. \ref {Cornu} for $\phi ''_1=-8. \,
10^5 \, {\rm fs}^2$. Two examples obtained via a numerical
resolution of the Schr\"{o}dinger equation are given. They
correspond to the experimental data shown on the same figure. The
probability amplitude follows a double spiral starting from
$(0,0)$. Three regions can be distinguished. The two spirals
result from contributions before (I) and after (III) resonance.
The intermediate region (II : time interval $ \left[ {t_0 - \sqrt
{\phi_1 ''} ;\,t_0  + \sqrt {\phi_1 ''} } \right] $) corresponds
to the passage through resonance. It provides the main
contribution to the population. This contribution is
characterized by a sharp increase of $\left| a_e \left( t \right)
\right|$, in an almost "straight" direction as expected from the
stationary phase approximation applied to Eq. \ref{amplitude e}.
The two spirals play different roles. The first one (I) winds
round the origin with an increasing radius. The resulting
probability increases thus slowly and regularly. After resonance
(III), a second spiral winds round the asymptotic value leading
to strong oscillations. For a central laser frequency equal to
the resonance frequency, the two spirals have the same weight
(Fig. \ref {Cornu}a). Detuning of the laser frequency changes
their relative weight (Fig. \ref {Cornu}b). For negative chirp and
negative detuning, the second spiral (after resonance) has a
larger weight since resonance is reached during the first half of
the pulse. This leads to a higher number of oscillations, as can
be observed by comparing Fig. \ref {resonance} and \ref
{belowresonance}. Conversely, a negative chirp and positive
detuning reduces the number of oscillations.

The general expression of the excited state amplitude (Eq.
\ref{amplitude e}) presents strong similarities with Fresnel
diffraction of a gaussian beam by a sharp edge. Effectively, the
spirals displayed on Fig. \ref{Cornu} are similar to the
well-known Cornu spirals. In Eq. \ref{amplitude e} the variable
$t$ corresponds to the position of the sharp edge in the gaussian
beam. Without detuning, Eq. \ref{amplitude e} reproduces the
classical diffraction figure {\it on axis}. The effect of
detuning can be understood by observing in a slant direction
$\theta$, with $\sin \theta \propto \Delta\omega$.

Other analogies between time and space Fresnel diffraction have been
reported by several groups \cite{NoordamDiffract92,legouet00}. One
example deals with two-photon absorption (or frequency doubling) from a
chirped pulse  \cite{NoordamDiffract92}. The spatial coordinate was
associated to the laser frequency instead of the pump probe delay in
our present case. By inserting masks on the laser spectrum, the
equivalent of a Fresnel zone lens was reproduced in the spectral
domain. The frequency doubled pulse was thus spectrally focused. In our
case, pulse shaping of the laser pump pulse should provide similar
effects as a Fresnel lens. For instance, by slicing the temporal
profile of the pump pulse, it should be possible to suppress the
destructive interference contributions to $a_e \left( t \right)$ and
increase significantly its asymptotic value.

We have presented in this letter direct measurements of Coherent
Transients observed on the excited state population at an
ultrashort scale. The present experiments have been performed with
a linear chirp. It provides an accurate measurement of the chirp.
This scheme can be extended to nonlinear variations of the chirp
which could thus be measured in a direct way. Moreover, in
pump-probe experiments as well as in any experiment based on a
combination of ultrashort pulses, such coherent transients should
be taken carefully into account to analyze the data. Finally
based on this phenomenon, new pulse shaping schemes could be
developed to improve the transfer efficiency.

We enjoyed fruitful discussions with Jacques Vigu\'{e}, François
Biraben, François Nez, Pierre Glorieux, Jacques Dupont-Roc and
Serge Haroche. Stimulating advices from E. Riedle and M.
Zavelani-Rossi are sincerely acknowledged.




\begin{thebibliography}{99}



\bibitem{AllenEberly74} L. Allen and J. H. Eberly, {\it Optical resonance and two-level atoms},  (Dover publications, New York, 1974).
\bibitem{Abragam61} A. Abragam, {\it Principles of Nuclear Magnetism},  (1961).
\bibitem{Judson92} R. S. Judson and H. Rabitz,  Phys. Rev. Lett. {\bf 68}, 1500 (1992).
\bibitem{Eberly85} J. Oreg, G. Kazak and J. H. Eberly,  Phys. Rev. A {\bf 32}, 2776 (1985).
\bibitem{CorkumChirp90} S. Chelkowski, A. D. Bandrauk and P. B. Corkum,  Phys. Rev. Lett. {\bf 65}, 2355 (1990).
\bibitem{WarrenI291} J. S. Melinger, A. Hariharan, S. R. Gandhi and W. S. Warren,  J. Chem. Phys. {\bf 95}, 2210 (1991).
\bibitem{Noordamladder92} B. Broers, H. B. van Linden van den Heuvell and L. D. Noordam,  Phys. Rev. Lett. {\bf 69}, 2062 (1992).
\bibitem{Band94b} Y. B. Band,  Phys. Rev. A {\bf 50}, 5046 (1994).
\bibitem{NoordamNO98} D. J. Maas, D. I. Duncan, R. B. Vrijen, W. J. van der Zande and L. D. Noordam,  Chem. Phys. Lett. {\bf 290}, 75 (1998).
\bibitem{Bergmann90} U. Gaubatz, P. Rudecki, S. Schiemann and K. Bergmann,  J. Chem. Phys. {\bf 92}, 5363 (1990).
\bibitem{Trushin00} S. A. Trushin, W. Fuss and W. E. Schmid,  Chem. Phys. {\bf 259}, 313 (2000).
\bibitem{Brewer71} R. G. Brewer and R. L. Shoemaker,  Phys. Rev. Lett. {\bf 27}, 631 (1971).
\bibitem{Glorieux72} B. Macke and P. Glorieux,  Chem. Phys. Lett. {\bf 14}, 85 (1972).
\bibitem{Grischkowsky76} D. Grischkowsky,  Opt. Commun. {\bf 18}, 69 (1976).
\bibitem{Grischkowsky86} J. E. Rothenberg and D. Grischkowsky,  J. Opt. Soc. Am. B {\bf 3}, 1235 (1986).
\bibitem{Vitanov99} N. V. Vitanov,  Phys. Rev. A {\bf 59}, 988 (1999).
\bibitem{DeSilvestri97} G. Cerullo, M. Nisoli and S. De Silvestri,  Applied-Physics-Letters {\bf 71}, 3616 (1997).
\bibitem{Jones95MultiEnhance} R. R. Jones,  Phys. Rev. Lett. {\bf 75}, 1491 (1995).
\bibitem{Zamith2000a} S. Zamith, M. A. Bouchene, E. Sokell, C. Nicole, V. Blanchet and B. Girard,  Eur. Phys. J. D {\bf 12}, 255 (2000).
\bibitem{NoordamDiffract92} B. Broers, L. D. Noordam and H. B. van Linden van den Heuvell,  Phys. Rev. A {\bf 46}, 2749 (1992).
\bibitem{legouet00} L. M\'enager, I. Lorger\'e and J. L. Legouet,  Opt. Lett. {\bf 25}, 1316 (2000).
\end{thebibliography}
\end{document}